# Molecule-based microelectromechanical sensors

Matias Urdampilleta,[1,2,*] Cedric Ayela,[3,4,5,*] Pierre-Henri Ducrot,[3,4,5] Daniel Rosario-Amorin,[1,2] Abhishake Mondal,[1,2] Mathieu Rouzières,[1,2] Pierre Dechambenoit,[1,2] Corine Mathonière,[6,7] Fabrice Mathieu,[8] Isabelle Dufour,[3,4,5] and Rodolphe Clérac[1,2,*]

[1] CNRS, CRPP, UPR 8641, 33600 Pessac (France). E-Mail: clerac@crpp-bordeaux.cnrs.fr (R.C.) and matias.urdampilleta@neel.cnrs.fr (M.U.)

[2] Univ. Bordeaux, CRPP, UPR 8641, 33600 Pessac (France).

[3] Univ. Bordeaux, IMS, UMR 5218, F-33405 Talence, (France). E-Mail: cedric.ayela@ims-bordeaux.fr

[4] CNRS, IMS, UMR 5218, F-33405 Talence, France.

[5] Bordeaux INP, IMS, UMR 5218, F-33405 Talence, France.

[6] CNRS, ICMCB, UPR 9048, 33608 Pessac Cedex (France).

[7] Univ. Bordeaux, ICMCB, UPR 9048, 33600 Pessac, (France).

[8] LAAS, CNRS et Université de Toulouse, INSA, UPS, F-31077 Toulouse, France

**Abstract: Incorporating functional molecules into sensor devices is an emerging area in molecular electronics that aims at exploiting the sensitivity of different molecules to their environment[1–4] and turning it into an electrical signal. Among the emergent and integrated sensors, microelectromechanical systems (MEMS) are promising for their extreme sensitivity to mechanical events[5]. However, to bring new functions to these devices, the functionalization of their surface with multifunctional molecules is required. Herein, we present original hybrid devices made of an organic microelectromechanical resonator functionalized with switchable magnetic molecules. The change of their mechanical properties and geometry induced by the switching of their magnetic state at a molecular level alters the device's dynamical behavior, resulting in a change of the resonance frequency. We demonstrate that these devices can be operated to sense light or thermal excitation. Moreover, thanks to the collective interaction of the switchable molecules, the device behaves as a non-volatile memory. Our results open up broad prospects of new flexible photo- and thermo-active hybrid devices for molecule-based data storage and sensors.**

The development of molecule-based electronic devices for sensing or computation requires the synthesis of new materials and the fabrication of new architectures. As a result, switchable molecules, which can be turned into two different states using an external stimulus, have been thoroughly investigated during the last decade. Their switching can be activated using a wide range of external excitations such as photons[6], tunneling electrons[7], chemical species[8], or temperature[9]. Among the large variety of molecular switches, magnetic molecules have attracted a lot of interest as they combine interesting magnetic, electronic and mechanical properties[10] and can be, in certain cases, photoactivated[11,12]. These magnetic molecular switches (MMSs)[13] possess two accessible states with high-spin (HS) and low-spin (LS) configurations. The switching between these two states can be induced either by a change in the spin configuration of a metal center (e.g. spin crossover compounds[14,15]) or by the transfer of one electron from one metal site to another (e.g. electron transfer molecules[16]). The HS/LS conversion is commonly driven by a competition between the enthalpy and the entropy of the system. Thus, a change of the molecular spin is observed as a function of the temperature with the LS state being favorable below the crossover temperature. Remarkably, once in their LS state, the MMSs can be photo-switched in their HS state[11,14].

Interestingly, the magnetic switching is associated by a drastic change of the molecule volume as schematized in Figure 1a, leading to a change of the mechanical properties and geometry of the material. Recently, it has been proposed to exploit these properties by integrating MMSs in a free-standing mechanical system[17–19]. An alternative experimental approach consists in using an inorganic microelectromechanical system (MEMS) functionalized with switchable magnetic molecules for which the change of mechanical properties affects the resonance frequency of the device[20]. Another interesting and powerful approach to expand the application capability of the molecular switches is the use of organic MEMS[21]. In this case, large variations of the organic MEMS resonance could be observed when a surface stress is generated by the volume change of the MMSs. This idea has been experimentally demonstrated here. We have measured unprecedentedly large resonance frequency shifts of the hybrid device when the molecules experience a switching, either induced by thermal or optical excitations. Moreover, the present results have been reproduced with different molecular systems under different forms, thus demonstrating the versatility of our approach.

The piezoelectric MEMS used in this work are organic and fabricated at low-cost thanks to a rapid and simple process[21]. They are made of a multilayer assembly consisting of a polyethylene naphtalate (PEN) substrate covered with the piezoelectric copolymer,

poly(vinylidene fluoride-trifluoroethylene) (PVDF-TrFE), sandwiched between two thin layers of aluminum[21] (Figures 1b and 1c). Thanks to the polymer piezoelectricity, the resonators are actuated by applying a voltage across the two parallel aluminum electrodes. The detection of the mechanical resonance is achieved by measuring the change of the motional impedance, $Z_m$, of the piezoelectric layer, induced by the displacement (Methods). Prior to the functionalization of the MEMS with MMSs, the temperature dependence of the first flexural mode, $f_0$, was investigated. For this purpose, the MEMS device was cooled down using a liquid nitrogen cryostat and the resonance curves were collected for different temperatures (Figure 1d). The characteristic frequency $f_0$ exhibits a quadratic behavior mainly due to the dependence of the Young's modulus with temperature.

To demonstrate the feasibility of our detection method, we focused first on the [Fe(dmbpy)(H$_2$B(pz)$_2$)$_2$] spin crossover compound (noted **SCO1**; dmbpy = 4,4′-dimethyl-2,2′-dipyridyl and H$_2$B(pz)$_2$ = dihydrobis(pyrazolyl)borate; Figures 1e and S1), as it exhibits a large change of unit cell volume (6 %) at the crossover temperature (See supporting information). **SCO1** was either dissolved in methanol and dropcasted onto the MEMS surface (device A; Figure 1e), or directly sublimated using a thermal evaporator (device B), as reported in reference 22. Figure 2a shows the resonance frequency for device A as a function of the temperature for pristine and hybrid resonators. A clear decrease of $f_0$ is observed around 165 K for the hybrid resonator, which corresponds to the spin crossover temperature measured by standard magnetometry on bulk samples (Figure S2). To highlight the effect of the spin crossover on the mechanical response, we apply a quadratic correction to $f_0$ which suppresses the intrinsic temperature dependence of the resonator structure. The Figure 2b presents the evolution of $\Delta f^*/f^*$, the relative variation of the corrected frequency that is directly compared to the signature of the spin crossover phenomenon obtained by direct magnetic measurements (Figures 2c and 2d for devices A and B, respectively). These results prove that the large variations of the organic MEMS resonance frequency are induced by the spin crossover phenomenon of the molecules. Moreover, it clearly indicates that the deposition method does not strongly affect the crossover.

The shift of the resonance frequency is the consequence of a mechanical alteration of the resonator triggered by the switchable molecules at the spin crossover temperature. Two main effects have been discussed in the literature[17,20]: (i) the change of mechanical properties of the material (Young's modulus, mass density) and (ii) the change of surface stress at the interface between MMSs and the resonator. The total relative frequency shift, $[\Delta f_0/f_0]^{Total}$, is thus given by Equation 1:

$$\left[\frac{\Delta f_0}{f_0}\right]^{Total} = \left[\frac{\Delta f_0}{f_0}\right]^{Material} + \left[\frac{\Delta f_0}{f_0}\right]^{Stress} \quad (1)$$

The relative frequency shift due to mechanical properties, $[\Delta f_0/f_0]^{Material}$, was discussed in reference[20]. Applying this model to our devices gives a theoretical $[\Delta f_0/f_0]^{Material}$ value of + 0.16% (See supporting information). This estimation is clearly contrasting with our results, which reveal a total shift of − 2.25% (Figures 2b). The detected resonance shift is thus dominated by the surface stress that implies two contributions[19]: (i) the differential surface stress, $\sigma_S^{Diff} = \sigma_S^{Upper} - \sigma_S^{Lower}$ (with $\sigma_S^{Upper}$ and $\sigma_S^{Lower}$, the surface stresses applied to the upper and lower faces of the cantilever) that induces a bending of the cantilever and (ii) the total surface stress: $\sigma_S^{Total} = \sigma_S^{Upper} + \sigma_S^{Lower}$. While the first component is known to have a negligible effect on the resonance frequency[23], the total surface stress strongly influences the cantilever resonance frequency[24]. The analytical expression of $[\Delta f_0/f_0]^{Stress}$ in relation with the total surface stress is explicitly given by Lachut and Sader[23], Equation 2:

$$\left[\frac{\Delta f_0}{f_0}\right]^{Stress} = -0.042 \frac{\nu(1-\nu)\sigma_S^{Total}}{Eh}\left(\frac{b}{L}\right)\left(\frac{b}{h}\right)^2 \quad (2)$$

with $L$ being the cantilever length, $b$ the width, $h$ the thickness, $E$ the equivalent Young's modulus of the multistacked device, and $\nu$ the Poisson ratio. Applying this approach to our hybrid MMS/MEMS resonators gives a stress induced in the MMSs material of 250 MPa ($\sigma_{MMS} = \frac{\sigma_S^{Total}}{h_{MMS}}$) for an experimental resonance frequency shift of − 2.25%. To further demonstrate the key role of the total surface stress on the resonance frequency, the resonator was functionalized with MMSs on both its top and bottom faces as a significant increase of the shift is expected, whereas an effect of the differential surface stress should cancel this shift. As shown in Figure 2b, $[\Delta f_0/f_0]^{Total}$ indeed increases to − 3.8%, in agreement with the prediction of our model. It is also possible to compare the surface stress value, $\sigma_{MMS}$, of 250 MPa obtained with the above model with the stress, $\sigma_{MMS}^* = \varepsilon_{MMS}^* E_{MMS}/(1 - \nu_{MMS})$, that should be induced by eigenstrain due to MMS volume change ($\varepsilon_{MMS}^* = \frac{1}{3}\Delta V/V$). The numerical estimation of $\sigma^*_{MMS}$ leads to 145 MPa, which is in relative good agreement with the $\sigma_{MMS}$ value obtained above, especially considering that $\sigma^*_{MMS}$ does not take into account effects at the interface between MMSs and the MEMS resonator. The present analysis establishes that the origin of the frequency shift observed is the total surface stress induced by the volume change of the switchable molecules. This work also underlines the benefits for molecular sensing using MMSs when combined with organic MEMS. Their low Young's modulus leads to highly sensitive sensors with large frequency shift responses.

In order to test the versatility of our molecule-based sensors, **SCO1** has been replaced by another switchable complex: Fe(MeOL-mCl)$_2$ (noted **SCO2**; MeOL-mCl: N'-((5-chloropyridin-2-yl)methylene)-4-methoxybenzohydrazonate; Figure S3; see supporting information). This compound exhibits a magnetic change from HS to LS through a first order phase transition induced by strong elastic interactions between molecules. As this transition is associated with a thermal hysteresis loop (Figure S4), this device allows us to confer to the sensor a memory effect. Similarly to device A, **SCO2** has been dropcasted onto the organic MEMS. From the temperature dependence of $f_0$, before and after deposition, the HS fraction of **SCO2** was estimated for a thermal cycle between 300 and 120 K (a cooling/warming cycle). As shown in Figure 3a, the spin crossover is preserved after deposition of **SCO2** and as expected for a spin transition, it appears more abrupt than for devices A and B coated with **SCO1**. The presence of a thermal hysteresis loop between 220 and 180 K, larger than in the **SCO2 bulk** material (Figure S4), is clearly observed. This reveals that the elastic interactions in this material are still active and even reinforced after the deposition. This result opens perspectives toward the use of spin transition materials in molecule-based sensors exhibiting non-volatile memory.

Another interesting property of some MMSs is their ability to be photo-switched between their different magnetic states. To probe the possibility to build photo-active sensors, an electron transfer molecule, ({[(pzTp)Fe(CN)$_3$]$_4$[Co(pz)$_3$CCH$_2$OH]$_4$[ClO$_4$]$_4$}·13DMF·4H$_2$O noted **ET3**; pzTp = tetrakis(pyrazolyl)borate and (pz)$_3$CCH$_2$OH = 2,2,2-tris(1-pyrazolyl)ethanol; see supporting information) has been considered in this study as it can be photoswitched at liquid nitrogen temperature[25]. These molecules have been directly dropcasted onto the organic MEMS. The high-spin fraction of **ET3** was estimated from $f_0$ upon decreasing the temperature from 300 to 80 K as shown by the blue symbols in Figure 3b. A magnetic switching of the **ET3** molecules from the HS to the LS states is observed around 230 K. This magnetic conversion seems to be less abrupt than in the original

**ET3** material[25] and slightly shifted of about 20 K to lower temperatures. At 80 K, the **ET3**-based hybrid device was irradiated with white light ($P = 1$ mW/cm$^2$) while the time evolution of $f_0$ was recorded (Figure 3c). In less than 5 minutes, $f_0$ increases from 4.57 kHz to a saturation value of 4.63 kHz. The HS fraction estimated from this $f_0$ variation evidences the complete photo-conversion of the **ET3** molecules at 80 K (Figure 3b). When increasing the temperature in the dark (the red symbols in Figure 3b), the hybrid MEMS recovers the resonance frequency as before irradiation at 150 K. This feature, also seen in the original **ET3** material[21] at about 200 K, corresponds to the relaxation of the photo-induced HS molecules into their LS ground state; in other words, the system at 150 K has enough thermal energy to overcome the barrier separating the metastable and thermodynamic states. Further increase of the temperature leads to a recovery of the HS state above 220 K (Figure 3b). It is worth noting that the difference between the cooling and warming branches contrasts with the absence of thermal hysteresis in the bulk **ET3** material. This observation and the fact that the characteristic relaxation temperature of the photo-induced state is different for the hybrid device (150 K) and the original **ET3** compound (200 K) strongly suggests that the **ET3** molecules do not organize on the surface of the resonator with the same interaction pattern as in the bulk material. This last example highlights the remarkable versatility of these molecule-based microelectromechanical sensors: they can be triggered by temperature, as observed for the three reported hybrid devices, but also by light at a fixed temperature when functionalized with **ET3** molecules.

Combining organic MEMS with switchable magnetic molecules has allowed us to detect large variations of the resonance frequency induced by a small change of volume at the molecular scale. Providing that these molecules are extremely sensitive to external stimuli (such as temperature, light, pressure, chemical atmosphere…), these hybrid devices are promising systems for sensing applications. In this work, we have indeed demonstrated that the reported molecule-based MEMS could be operated as a temperature and light detector as well as a non-volatile memory. These results pave the way toward an integration of organic MEMS devices in molecular electronics.

**Methods:**
**Fabrication of devices**: The fabrication process starts with a 25 $\mu$m thick PEN film, used as a substrate, cleaned with isopropanol. Then, 300 nm of aluminum is evaporated through a PET (polyethylene terephthalate) shadow mask to pattern the bottom electrodes (for the free-standing cantilever and the reference capacitance). The deposition of the PVDF-TrFE layer is then achieved by using PVDF-TrFE powder (75-25 % in mole; Piezotech) dissolved in 2-butanone with a mass content of 20 %. This solution is spin-coated at 3500 rpm with a ramp of 1 second for 45 seconds, giving a thickness of about 4 $\mu$m. The resulting assembly is then annealed at 50 °C for 10 minutes to evaporate the solvent and at 140 °C for a hour to improve crystallinity. The aluminum top electrode is subsequently evaporated in the same conditions as the bottom one through a PET shadow mask. A 6 $\mu$m protective PDMS (polydimethylsiloxane; Sylgard 184, Dow Corning) layer is spin-coated on the top surface of the devices and cured at 80 °C for 2 hours. To finish the assembly process, the shape of the cantilever is obtained simply by xurography, thanks to a vinyl cutting machine Craft RoboPro CE6000 (Graftec Craft ROBO Pro). For proper use, the resulting MEMS resonators are glued on a glass blade with a double-sided adhesive tape, leaving the cantilever part suspended and the reference capacitance fixed. To induce piezoelectricity in the PVDF-TrFE film, the set-up is poled with a DC electric field of 100 V/$\mu$m for 10 minutes.

**Device measurement:** The total impedance $Z_t$ between the two aluminum electrodes is composed of the motional impedance $Z_m$ with a constant capacitance $C_0$ in parallel ($C_0$ corresponds to the geometrical capacitor made of the PVDF-TrFE dielectric layer sandwiched between the two aluminum electrodes). To measure directly the variation of $Z_m$, a reference capacitance $C_0$ has been integrated on the MEMS chip, see Figure 1b. It is composed of the same stacked layers but not free to move. An alternative signal is then applied on both structures with an independent, tunable amplitude of opposite sign to compensate the effect of $C_0$. The voltage at the common electrode is then zero except when the free structure starts to move. Both the amplitude and phase are obtained using an IQ demodulator and then the resonance frequency $f_0$ can easily be measured by a dedicated electronic card.

**Synthesis and characterization of [Fe(dmbpy)(H$_2$B(pz)$_2$)$_2$] (SCO1):** To a solution of KH$_2$B(pz)$_2$ (200 mg, 1.07 mmol) in MeOH (4 mL) was added Fe(ClO$_4$)$_2$•6H$_2$O (195 mg, 0.54 mmol). The mixture was stirred for 10 minutes and the precipitate of potassium perchlorate was removed by filtration. A solution of 4,4'-dimethyl-2,2'-bipyridine (99 mg, 0.54 mmol) in MeOH (2 mL) was added dropwise to the colorless methanolic solution of [Fe(H$_2$B(pz)$_2$)$_2$] and the mixture was allowed to stir for one hour. After filtration, the dark violet powder is dissolved in a CH$_2$Cl$_2$/MeOH mixture (9 mL, 2:1) and the clear solution is slowly evaporated. After one week, the resulting violet needle-shape single crystals were filtered and washed with cold methanol: Yield 185 mg (64 %). Elemental analysis Calc. C$_{24}$H$_{28}$N$_{10}$FeB$_2$: C, 53.97 (53.56); H, 5.28 (5.29); N, 26.23 (26.21). Selected FT-IR data (ATR, cm$^{-1}$): 2394 (s), 2288 (s), 1601 (s), 1558 (w), 1498 (m), 1397 (s), 1201 (m), 1155 (vs), 1051

(s), 1014 (w), 974 (s), 877 (s), 823 (s), 771 (vs), 717 (m), 639 (s). Crystallographic data are given in Table S1 as well as an ORTEP view of **SCO1** at 100 K in Figure S1. Magnetic properties of a bulk **SCO1** sample are shown in Figure S2.

**Synthesis and characterization of Fe(MeOL-mCl)$_2$ (SCO2):** N'-((5-chloropyridin-2-yl)methylene)-4-methoxybenzohydrazide (100 mg, 0.34 mmol), trimethylamine (100 μL, 1.36 mmol) and Fe(ClO$_4$)$_2$•6H$_2$O (62 mg, 0.17 mmol) were combined in methanol (8 mL) and the mixture was stirred at room temperature for 1 hour. The resulting olive green powder of [Fe(MeOL-mCl)$_2$] was collected by filtration and washed with methanol: Yield 89 mg (81 %). Elemental analysis Calc. (Found) for C$_{28}$H$_{22}$Cl$_2$FeN$_6$O$_4$: C, 53.11 (52.32); N, 13.27 (12.98), H, 3.50 (3.69). Selected FT-IR data (ATR, cm$^{-1}$): 1600 (s), 1577 (s), 1508 (m), 1447 (s), 1405 (m), 1335 (vs), 1302 (s), 1287 (s), 1248 (vs), 1159 (vs), 1141 (m), 1120 (m), 1099 (m), 1059 (s), 1022 (s), 917 (m), 906 (m), 865 (m), 839 (s), 764 (vs), 697 (s), 622 (s). Single-crystals of [Fe(MeOL-mCl)$_2$] were obtained by slow diffusion of MeOH (6 mL) into a solution of complex in CH$_2$Cl$_2$ (1.5 mL). Crystallographic data are given in Table S2 as well as an ORTEP view of **SCO2** at 120 K in Figure S3. Magnetic properties of a bulk **SCO2** sample are shown in Figure S4.

**Synthesis and characterization of {[(pzTp)Fe(CN)$_3$]$_4$[Co(pz)$_3$CCH$_2$OH]$_4$[ClO$_4$]$_4$}·13DMF·4H$_2$O (ET3):** The synthesis was adapted from the one published in reference 25. Treatment of Co(ClO$_4$)$_2$•6H$_2$O (0.366 g, 1.00 mmol) in DMF (10 mL) with [NBu$_4$][(pzTp)Fe$^{III}$(CN)$_3$] (pzTp = tetrakis(pyrazolyl)borate 0.660 g, 1.00 mmol) afforded a dark red solution that was magnetically stirred for two hours. Addition of Et$_2$O (60 mL) precipitated a red oil and the supernatant was decanted; the red oil was washed with Et$_2$O (20 mL) and evacuated to dryness affording a red powder. The red solid was extracted into CH$_2$Cl$_2$ (15 mL) and filtered; (pz)$_3$CCH$_2$OH (0.244 g, 1.0 mmol; (pz)$_3$CCH$_2$OH = 2,2,2-tris(*1*-pyrazolyl*)ethanol) was added, and the red mixture was allowed to stir for 2 hours. The solution was evacuated to dryness at room temperature. The red residue was dissolved into DMF (8 mL) and divided into two parts. Each part was layered with Et$_2$O (15 mL), and red block type crystals were collected after 4 days. Yield: 0.65 g (61 %). All the physical characterizations (IR, X-ray structure, magnetic properties and EA) were found to be identical to the data reported in reference 25. It is worth mentioning that the cyanido stretching absorption observed by IR spectroscopy for **ET3** (2168 cm$^{-1}$) remains almost the same after depositing the compound on the MEMS device (2164 cm$^{-1}$). A view of the **ET3** molecular structure is shown in Figure S5.

**Acknowledgments:** This study has been carried out with financial support from the Institut de Chimie of the Centre National de la Recherche Scientifique (CNRS), the Université de Bordeaux, the Région Aquitaine, the GDR MCM2, the Agence Nationale de la Recherche (ANR; N°ANR-13-PDOC-0032) and with anecdotal funds and support (2 months salary for M.U.) from the LabEx AMADEus (ANR-10-LABX-42) in the framework of IdEx Bordeaux (ANR-10-IDEX-03-02), that is, the Investissements d'Avenir programme of the French government managed by the ANR. We also thank M. Abbas and L. Hirsch for experimental support, and J. F. Tassin who has encouraged us to develop this project and to have facilitated the short stay of M.U. at the Centre de Recherche Paul Pascal.

**Author contributions:** M.U., C.A. and R.C. initiated the project and designed the experiments. P.-H.D. and C.A fabricated the MEMS devices. D.R.A., A.M., C.M., P.D. and R.C. synthesized and characterized the MMSs. F.M. conceived the electronic board for the MEMS excitation and data acquisition. The experiments were carried out by M.U., M.R. and C.A.; I.D. and C.A. developed the model. All authors discussed the results. M.U., R.C. and C.A. wrote the manuscript with inputs from all authors.

**Additional information**: Correspondence and request for materials should be addressed to M.U., R.C. or C.A.

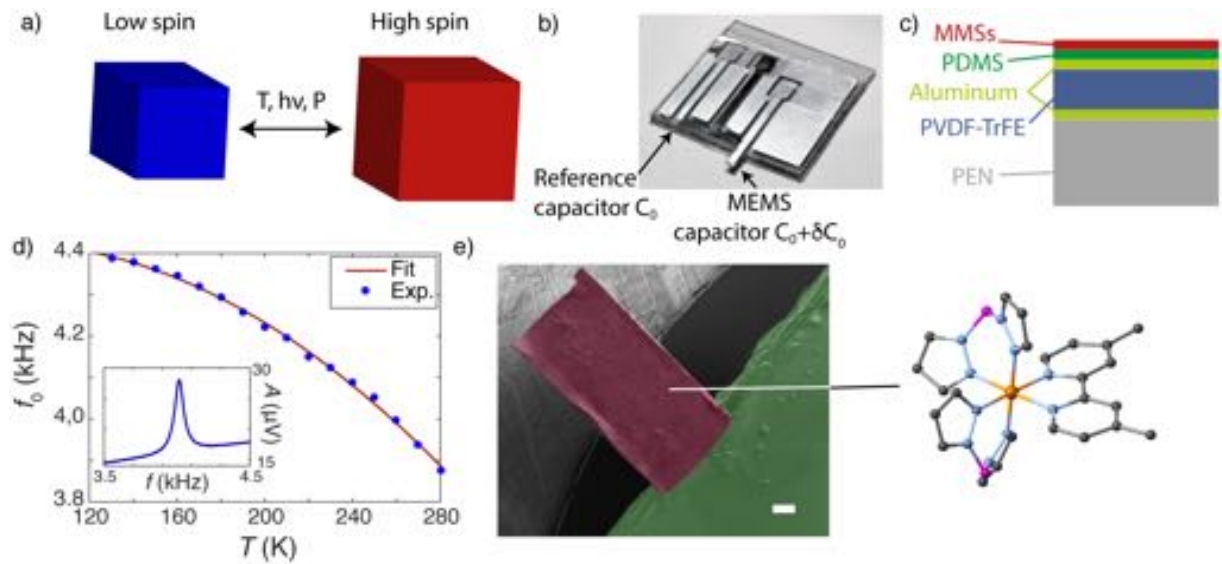

*Figure 1 :* A molecule-based MEMS : (a) Scheme illustrating that the magnetic switching can be activated in a MMS using different stimuli: temperature, light or pressure. The change of magnetic state is accompanied by a change of the molecule volume. (b) View of a chip implemented with the organic piezoelectric microelectromechanical system and the reference structure. (c) Scheme of the cut view of the MEMS suspended part highlighting its layered structure: the PEN substrate is covered with a piezoelectric PVDF-TrFE layer sandwiched between two aluminum electrodes. A protective PDMS layer is deposited on the top of the structure and covered with the switchable magnetic molecule. (d) Temperature dependence of the resonance frequency, $f_0$, of the pristine structure. The stiffness of the resonator increases with decreasing temperature leading to an increase of the resonance frequency when lowering the temperature. Inset: a typical resonance spectrum of the well-compensated piezoelectric resonator. (e) SEM micrograph of the molecule-based MEMS with the functionalized area in bordeaux that contrasts with the uncovered area in green (artificially colorized view). Scale bar is 100 µm. Inset: molecular structure of complex **SCO1**, that exhibits spin-crossover (SCO) properties.

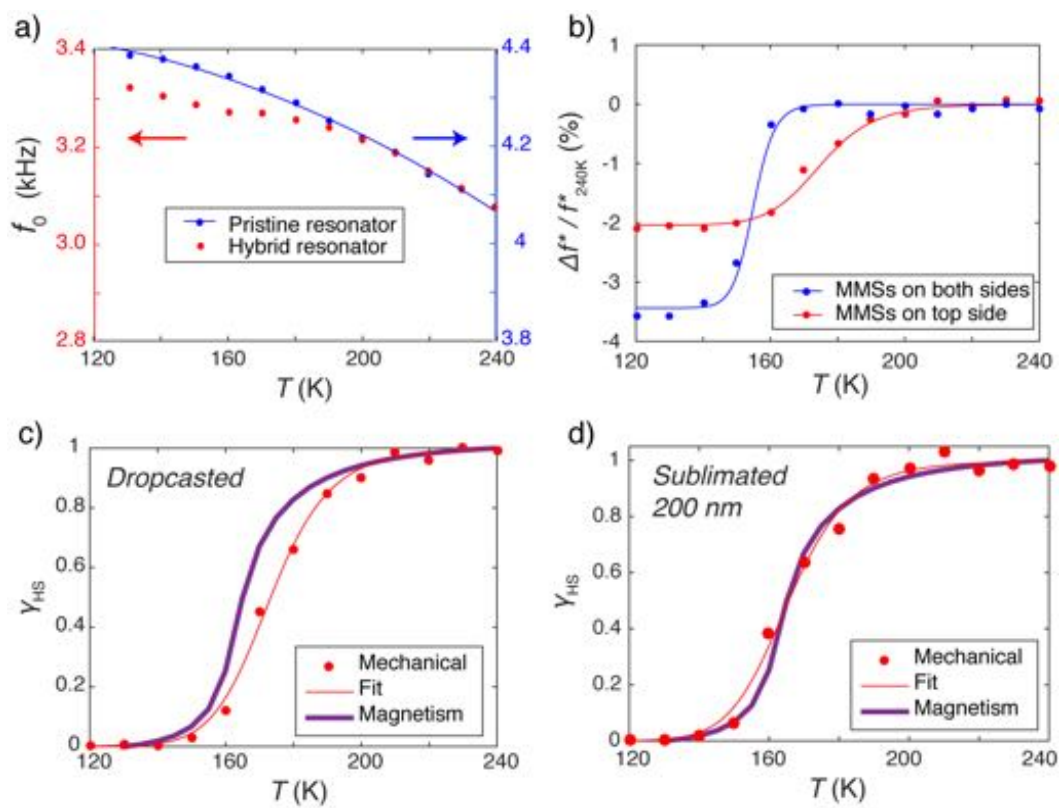

*Figure 2:* Molecule-based MEMS and detection of the spin crossover phenomenon: (a) Comparison of the resonance frequency as a function of temperature before and after functionalization of the resonator (here as an illustrating example with **SCO1** molecules). The hybrid resonator shows a drop of the resonance frequency around 165 K. (b) Relative variation of the resonance frequency corrected with the quadratic behavior of the pristine resonator. The red (blue) curve corresponds to a single (double)-side functionalization by MMSs (dropcasted). (c & d) High spin fraction of the **SCO1** molecules as a function of the temperature. The red dots are extracted from the mechanical response (resonance frequency) of the resonator functionalized with dropcasted (b; Device A; frequency data are shown in Figure 2a) or sublimated (c; Device B) MMSs. The experimental data are then fitted using a Fermi distribution (continuous line). The bordeaux curve corresponds to magnetometry measurement performed on a polycrystalline sample of **SCO1** using a SQUID magnetometer (See Figure S2).

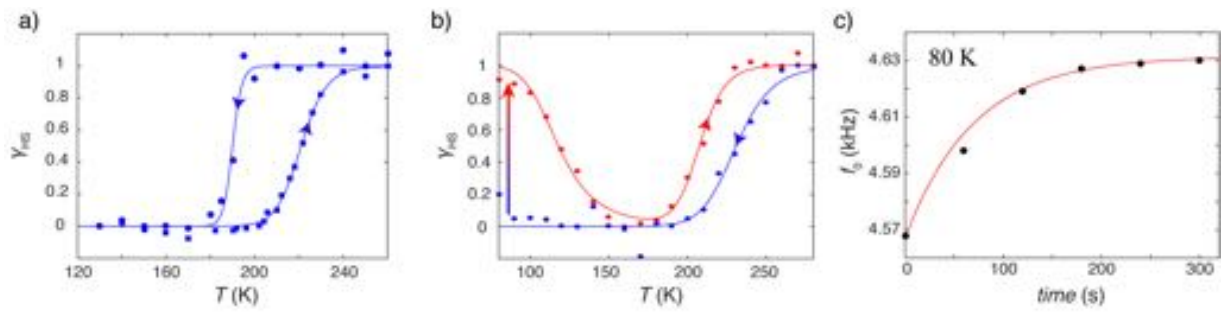

*Figure 3:* Bistable molecule-based MEMS: (a) High spin fraction of **SCO2** molecules as a function of the temperature. The blue dots and squares are deduced from the experimental mechanical response (resonance frequency) of the organic resonator functionalized with dropcasted MMSs in cooling and heating modes, respectively. The continuous lines are guides for the eyes. The first order phase transition of the deposited molecule-based material induces a thermal hysteretic effect giving rise to a bistable behavior. (b) High spin fraction of **ET3** molecules as a function of the temperature. The blue and red dots are deduced from the experimental mechanical response (resonance frequency) of the organic resonator functionalized with dropcasted MMSs in cooling (in the dark) and heating (in the dark after light irradiation at 80 K) modes, respectively. The continuous lines are guides for the eyes. (c) Time evolution of the resonance frequency of the organic MEMS functionalized with **ET3** during illumination at 80 K with white light ($P = 1$ mW/cm$^2$).

# Supporting Online Material

**Additional experimental information.**
**Magnetic properties:** The magnetic measurements were carried out with the use of Quantum Design MPMS-XL SQUID magnetometer. These instruments work between 1.8 and 400 K with applied dc fields ranging from − 7 to 7 T. Measurements were performed on a polycrystalline samples of **SCO1** (12.95 mg) and **SCO2** (18.92 mg) sealed in a polyethylene bag (3 × 0.5 × 0.02 cm; typical 20 to 40 mg). Prior to the experiments, the field-dependent magnetization was measured at 100 K in order to confirm the absence of any bulk ferromagnetic impurities. The magnetic data were corrected for the sample holder and the intrinsic diamagnetic contributions.

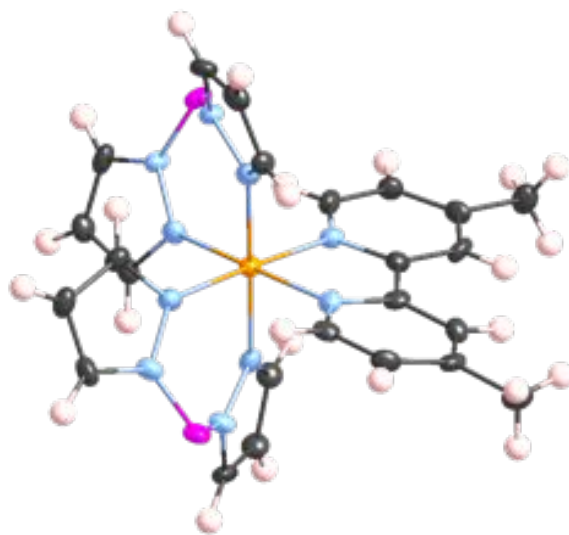

*Figure S1.* ORTEP view of [Fe(dmbpy)(H$_2$B(pz)$_2$)$_2$] (**SCO1**) at 100 K. Thermal ellipsoids are depicted at 50 % probability. Fe orange, C grey, N blue, B pink, H light pink.

*Table S1.* Crystallographic data for [Fe(dmbpy)(H$_2$B(Pz)$_2$)$_2$] (**SCO1**).

| Temperature, K | 100 (low-spin) | 250 (high-spin) |
|---|---|---|
| Crystal description | Turquoise needle | Violet needle |
| Moiety formula | C$_{24}$H$_{28}$FeB$_2$N$_{10}$ | C$_{24}$H$_{28}$FeB$_2$N$_{10}$ |
| Empirical formula | C$_{24}$H$_{28}$FeB$_2$N$_{10}$ | C$_{24}$H$_{28}$FeB$_2$N$_{10}$ |
| Formula weight | 534.01 | 534.01 |
| Crystal system | Monoclinic | Monoclinic |
| Space group | *P2$_1$/c* | *P2$_1$/c* |
| Wavelength, Å | 0.71073 | 0.71073 |
| *a*, Å | 11.0414(10) | 11.0705(17) |
| *b*, Å | 14.2520(15) | 14.626(3) |
| *c*, Å | 19.3138(14) | 19.889(3) |
| *β*, ° | 123.633(4) | 123.379(8) |
| *V*, Å$^3$ | 2530.5(4) | 2689.2(8) |
| Z | 4 | 4 |
| $\rho_{calcd}$, g.cm$^{-3}$ | 1.402 | 1.319 |
| $\mu_{MoK\alpha}$, mm$^{-1}$ | 0.631 | 0.594 |
| $R_1$[a] | 0.0503 | 0.0427 |
| $wR_2$[b] | 0.1087 | 0.1049 |
| *GoF*[c] | 1.024 | 1.043 |
| Fe-N$_{av}$, Å | 1.996 | 2.186 |
| $\Sigma_{(N-Fe-N)}$,[d] ° | 38.21 | 48.2 |

[a] $I > 2\sigma$ $R_1 = \Sigma ||F_o| - |F_c||/\Sigma |F_o|$, [b] $wR_2 = [\Sigma w(F_o^2 - F_c^2)^2/\Sigma w(F_o^2)^2]^{1/2}$, [c] *GoF* (goodness of fit on $F^2$) = $\{\Sigma [w(F_o^2 - F_c^2)^2]/(n-p)\}^{1/2}$, where n is the number of reflections and p is the total number of refined parameters. [d] $\Sigma = \Sigma_{i=1}^{12}|90 - \varphi_i|$, where $\varphi_i$ are cis N-Fe-N bond angles.

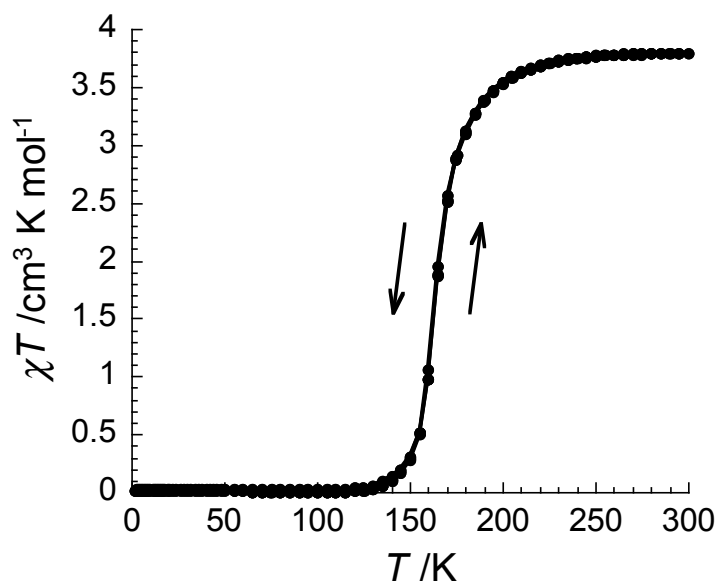

*Figure S2.* Temperature dependence of the $\chi T$ product ($\chi$ is defined as magnetic susceptibility equal to *M/H* per mole of [Fe(dmbpy)(H$_2$B(Pz)$_2$)$_2$], **SCO1** and *T* the temperature). Data collected at 0.1 and 1 T are identical in cooling and heating mode respectively (no significant thermal hysteresis effect), and thus they have been superposed on the figure. The solid lines are guides for the eyes.

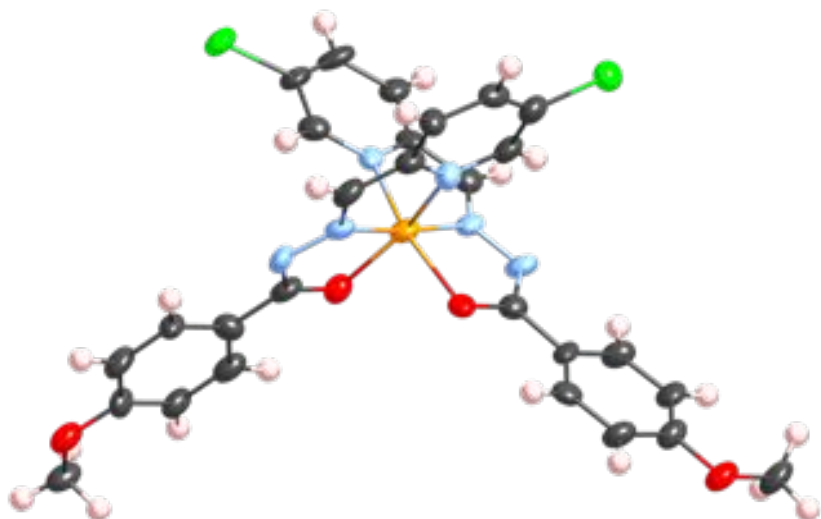

*Figure S3.* ORTEP view of [Fe(MeOL-mCl)₂] (**SCO2**) at 120 K. Thermal ellipsoids are depicted at 50 % probability. Fe orange, C grey, N blue, O red, Cl green, H light pink.

*Table S2.* Crystallographic data for [Fe(MeOL-mCl)$_2$] (**SCO2**).

| Temperature, K | 120 (low-spin) | 250 (high-spin) |
|---|---|---|
| **Crystal description** | Dark green plate | Dark green plate |
| **Moiety formula** | C$_{28}$H$_{22}$Cl$_2$FeN$_6$O$_4$ | C$_{28}$H$_{22}$Cl$_2$FeN$_6$O$_4$ |
| **Empirical formula** | C$_{28}$H$_{22}$Cl$_2$FeN$_6$O$_4$ | C$_{28}$H$_{22}$Cl$_2$FeN$_6$O$_4$ |
| **Formula weight** | 633.26 | 633.26 |
| **Crystal system** | Triclinic | Triclinic |
| **Space group** | *P*-1 | *P*-1 |
| **Wavelength, Å** | 0.71073 | 0.71073 |
| *a*, Å | 10.9324(19) | 11.2203(15) |
| *b*, Å | 11.4057(19) | 11.0521(14) |
| *c*, Å | 12.434(2) | 12.4511(18) |
| *α*, ° | 106.379(8) | 102.940(6) |
| *β*, ° | 112.323(6) | 113.914(5) |
| *γ*, ° | 96.457(8) | 94.326(5) |
| *V*, Å$^3$ | 1333.2(4) | 1351.5(3) |
| *Z* | 2 | 2 |
| *ρ*$_{calcd}$, g·cm$^{-3}$ | 1.577 | 1.556 |
| *μ*$_{MoKa}$, mm$^{-1}$ | 0.814 | 0.803 |
| *R*$_1$[a] | 0.0635 | 0.0549 |
| *wR*$_2$[b] | 0.1523 | 0.1337 |
| *GoF*[c] | 1.003 | 1.015 |
| Fe-N$_{av}$, Å | 1.910 | 2.163 |
| Fe-O$_{av}$, Å | 2.012 | 2.075 |
| Σ$_{(X-Fe-X)}$,[d] ° | 96.5 | 163.8 |

[a] $I > 2\sigma$ $R_1 = \Sigma ||F_o| - |F_c||/\Sigma |F_o|$, [b] $wR_2 = [\Sigma w(F_o^2 - F_c^2)^2/\Sigma w(F_o^2)^2]^{1/2}$, [c] *GoF* (goodness of fit on $F^2$) = $\{\Sigma[w(F_o^2 - F_c^2)^2]/(n-p)\}^{1/2}$, where n is the number of reflections and p is the total number of refined parameters. [d] $\Sigma = \Sigma_{i=1}^{12}|90 - \varphi_i|$, where $\varphi_i$ are cis X-Fe-X bond angles.

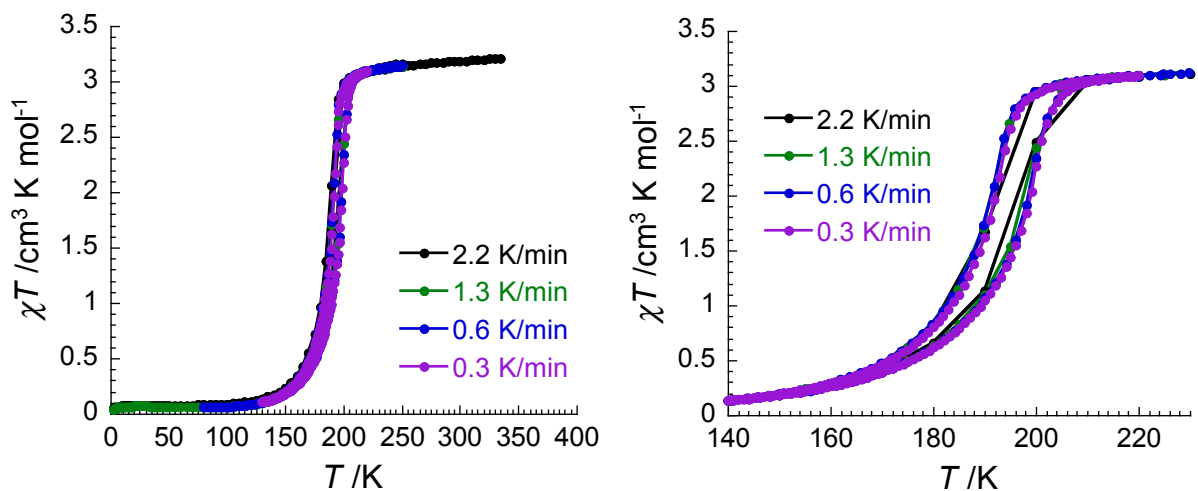

*Figure S4.* Temperature dependence of the $\chi T$ product ($\chi$ is defined as the magnetic susceptibility equal to $M/H$ per mole of [Fe(MeOL-mCl)$_2$], **SCO2**). Data collected at 0.1 or 1 T are identical, and thus they have been superposed on the figure. The solid lines are guides for the eyes.

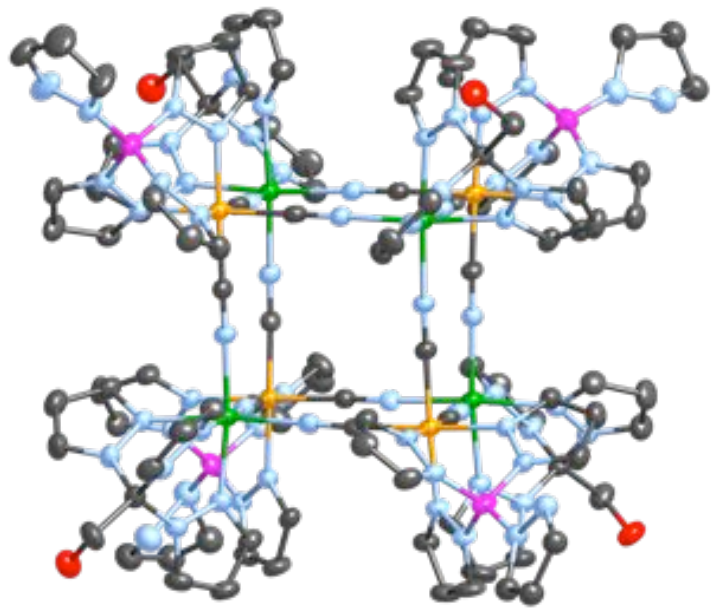

*Figure S5.* ORTEP view of {[(pzTp)Fe$^{III}$(CN)$_3$]$_4$[Co$^{II}$(pz)$_3$CCH$_2$OH]$_4$[ClO$_4$]$_4$}·**13DMF·4H$_2$O (ET3)** at 90 K from reference 25. Thermal ellipsoids are depicted at 20 % probability. Fe orange, Co green, C grey, N blue, O red, B pink. The magnetic properties of **ET3** can be found in reference 25.

**Detailed calculation of $[\Delta f_0/f_0]^{Material}$.**

The table below gathers the different mechanical parameters, which have been used in the high spin and low spin state for a very similar compound to **SCO1**[20]:

|  | High Spin | Low Spin | Relative variation |
|---|---|---|---|
| Young modulus (GPa) | 4.77 | 5.16 | + 8.2 % |
| Mass density (kg/m$^3$) | 1398 | 1471 | + 5.2 % |
| Poisson | 0.3 | 0.3 | 0.0 % |
| Volume (nm$^3$) | 2689.2 | 2530.5 | – 5.9 % |

The parameter variations at the magnetic switching change the mechanical behavior of the piezoelectric resonator as $f_0 = \frac{\lambda^2 h}{2\pi L^2}\sqrt{\frac{E}{12\rho}}$, where $\lambda$ is the eigenvalue (1.875) of the first mode, $h$ is the thickness of the resonator, $L$ its length, $E$ the Young's modulus and $\rho$ the mass density. The following device parameters need to be taken into considerations:

| Length $L$ (m) | 2×10$^{-3}$ |
|---|---|
| Width $b$ (m) | 1×10$^{-3}$ |
| Height $h$ (m) | 35.6×10$^{-6}$ |
| Thickness of SCO $h_{SCO}$ (m) | 4×10$^{-6}$ |
| **Equivalent Young modulus, $E_{eq}$ (Pa)** | **4.9×10$^9$** |
| **Equivalent mass density, $\rho_{eq}$ (kg/m$^3$)** | **1365** |
| Poisson constant, $\nu$ | 0.4 |

The above equivalent Young's modulus and mass density correspond to the whole layered structure (PEN+Al+PVDF-TrFE+PDMS) using:

|  | Thickness (m) | Mass density (kg/m$^3$) | Young's modulus (Pa) |
|---|---|---|---|
| PEN | 25×10$^{-6}$ | 1360 | 5×10$^9$ |
| Al | 0.6×10$^{-6}$ | 2700 | 69×10$^9$ |
| PVDF-TrFE | 4×10$^{-6}$ | 1800 | 2×10$^9$ |
| PDMS | 6×10$^{-6}$ | 965 | 1.8×10$^6$ |

and the following equations:

$$\rho_{eq} = \frac{\sum_i \rho_i h_i}{\sum_i h_i} \quad \text{and} \quad E_{eq} = \frac{4\pi^2 f_0^2 L^4 \times 12 \rho_{eq}}{\lambda_0^4 h^2}$$

Finally, when we take into account the change of Young's modulus and mass density on the resonance frequency at the spin crossover temperature, we obtain:[19]

$$\left[\frac{\Delta f_0}{f_0}\right]^{Material} = +0.16\ \%$$